\documentclass{Interspeech2024}
\usepackage{tablefootnote}
\usepackage{makecell}
\usepackage{multirow}

\interspeechcameraready

\title{Improvement Speaker Similarity for Zero-Shot Any-to-Any Voice Conversion of Whispered and Regular Speech}

\name[affiliation={1}]{Anastasia}{Avdeeva}
\name[affiliation={1}]{Aleksei}{Gusev}

\address{
  $^1$FluentaAI, Wilmington, USA}
\email{anastasia.avdeeva@fluenta.ai, aleksei.gusev@fluenta.ai}


\begin{document}

\maketitle

\begin{abstract}

Zero-shot voice conversion aims to transfer the voice of a source speaker to that of a speaker unseen during training, while preserving the content information. Although various methods have been proposed to reconstruct speaker information in generated speech, there is still room for improvement in achieving high similarity between generated and ground truth recordings. Furthermore, zero-shot voice conversion for speech in specific domains, such as whispered, remains an unexplored area. To address this problem, we propose a SpeakerVC model that can effectively perform zero-shot speech conversion in both voiced and whispered domains, while being lightweight and capable of running in streaming mode without significant quality degradation. In addition, we explore methods to improve the quality of speaker identity transfer and demonstrate their effectiveness for a variety of voice conversion systems.

\end{abstract}

\section{Introduction}

Voice Conversion (VC) is a task to convert a source speaker's voice to another one through modifying different voice characteristics such as speaker identity, accent and emotion while keeping linguistic information unchanged. Zero-shot VC is a more challenging task designed to effectively work with speakers unseen during training. Despite the significant success in the development of zero-shot VC systems, particularly with the adaptation of powerful Large Language Models (LLMs) for the VC task, and the use of neural audio codecs, a gap still exists between generated and authentic voices. Furthermore, zero-shot VC remains an unexplored area for various specific domains, such as whisper-to-speech VC. Whispered speech can be used by people with speech disorders and is also acknowledged as a technique to overcome stuttering \cite{whisper_works2}. Thus, converting whispered speech into regular speech creates various opportunities for speech interactions. However, the acoustic-phonetic distinctions between whispered and regular speech lead to degradation of existing VC systems when applied to whispered data \cite{our_icassp}. Also, although some previous works on whisper-to-speech conversion have been described as speaker-independent \cite{gan_whisper2speech} or zero-shot \cite{wesper}, none of these works have demonstrated speaker similarity evaluations. Thus, the proposals presented in this paper are as follows.

\begin{itemize}
    \item We propose zero-shot VC system based on StyleTTS2 \cite{TTS_style_tts} approach, called SpeakerVC, and demonstrate the ability of this system to work effectively within both a regular and whispered speech domains.
    \item We demonstrate that the proposed SpeakerVC system produces high-quality speaker reconstruction, comparable to current state-of-the-art (SOTA) zero-shot approaches, while also being lightweight and capable of running in streaming mode without significant quality degradation.
    \item We demonstrate that incorporating an additional speaker loss during training and increasing number of speakers in training dataset significantly improves the speaker similarity quality of the final VC system. Additionally, we show effectiveness of such an approach across various VC architectures. 
\end{itemize}

Samples from our proposed VC systems and publicly available systems can be found on the demo page\footnote{https://speakervc.github.io}.

\section{Related works}

This section briefly describes different methods for speaker identity reconstruction in zero-shot scenario within Text-to-Speech (TTS) and VC systems. Typically, the main idea of zero-shot approaches is to disentangle speaker and content information from the source and target speech and then reconstruct the source content with the target voice, unseen during the training procedure. A lot of developed systems rely on pretrained speaker models and use extracted speaker embeddings for voice reconstruction \cite{TTS_yourTTS} or jointly train speaker encoder with the rest of the pipeline \cite{VC_se_consistency_loss, VC_controlVC, VC_freeVC}. Some studies introduce a concept of a speaker consistency loss which aims to make embeddings from reference and reconstructed speech closer. This idea is successfully used both with pretrained speaker encoders \cite{TTS_yourTTS, VC_cycle_cons} and during joint training \cite{TTS_speaker_cons}. Also, various alternative methods for incorporating speaker or style information into models has been studied recently. For example, the AdaIN approach proposed in \cite{ada_in} can be effectively used for style transfer in the VC task \cite{VC_ada_in}. Other researches focus on improving the disentanglement between speaker and content information \cite{VC_speaker_erase}, demonstrating the importance of removing speaker information from linguistic content. This is achieved through methods such as quantization \cite{VC_speaker_dis} or instance normalization \cite{VC_ada_in, VC_auto_cycle_vc}. In contrast, the speaker embedding free system proposed in \cite{VC_sef_vc} relies on a position-agnostic cross-attention mechanism \cite{unicats}. Authors demonstrate its superiority over speaker embeddings-based systems. 

Special attention should be given to novel techniques for TTS and VC tasks, which utilize neural audio codecs \cite{encodec} and LLMs based models to process audio in the discrete domain. These techniques are applied to large training datasets and achieve excellent voice conversion quality. VALL-E \cite{vall_e} introduced an approach that treats TTS as a language model task and uses discrete audio codec codes as an intermediate representation. While this approach had shown good performance on the zero-shot TTS task, the Voicebox \cite{voicebox} generative model subsequently demonstrated superior quality. Despite their excellent quality, these models have serious limitations in terms of inference speed, which is an important aspect, especially for streaming systems. Thus, our aim in this research is to show that lightweight and faster systems can achieve comparable quality.

\section{Datasets}

This section describes datasets utilized for both training and evaluation. Information regarding the training datasets is summarized in Table \ref{tab:training_data}. Most of our experiments are performed using the VCTK + LibriTTS. For experiments on the extended dataset (D\_EXT) all training datasets were used. For VoxTube and TED\_X we use parts of the original datasets, additionally filtering out data with Signal-to-Noise Ratio less than 10dB, short speech duration and languages other than English. As these datasets do not contain any whispered speech samples, all data is converted to whisper with the Praat Toolkit\footnote{https://www.praatvocaltoolkit.com/whisper.html} before encoder feature extraction. 
% We use the original audio files, resampling them to 16kHz, 22.05kHz or 24kHz if necessary.

\begin{table}[th]
  \renewcommand*{\arraystretch}{1.1}
  \caption{Training datasets. For TED\_X, we assume that each video corresponds to one unique speaker and that the audio domain is between reading and spontaneous speech.}
  \label{tab:training_data}
  \centering
  \begin{tabular}{l|ccc}
    \toprule
    \textbf{Dataset} & \textbf{Dur. (hours)} & \textbf{Num. of spk} & \textbf{Domain} \\
    \midrule
    LibriTTS \cite{libri_tts} & 245 & 1151 & reading \\
    VCTK \cite{vctk} & 44 & 109 & reading \\
    VoxTube \cite{vox_tube} & 1454 & 1551 & spontan. \\
    TED\_X\tablefootnote{https://github.com/mauropelucchi/tedx\_dataset} & 309 & 2137 & -- \\
  \end{tabular}
\end{table}

%For testing purposes, we employ a variety of datasets to cover various domains and conditions. We use both voiced and whispered test corpora in our evaluation. 
For testing purposes, we use both voiced and whispered test corpora to cover various domains and conditions. Details regarding these datasets are provided in the Table \ref{tab:testing_data}. In the case of the Common Voice (CV) \cite{commonvoice} dataset we randomly select 100 speakers from the English portion of the dataset. Due to the absence of publicly available whispered spontaneous datasets, we collected the WhiSp dataset. The dataset was collected through the Upwork platform\footnote{https://www.upwork.com}. To compile the dataset, 40 English native speakers were asked to spontaneously respond to 40 questions using whispered voices. Additionally, we requested speakers to answer several questions using their regular speech to get the speaker's voice reference. 

\begin{table}[th]
  \renewcommand*{\arraystretch}{1.1}
  \caption{Testing datasets.}
  \label{tab:testing_data}
  \centering
  \begin{tabular}{l|cccc}
    \toprule
    \textbf{Dataset} & \textbf{\makecell{Dur. \\ (h.)}} & \textbf{\makecell{Num. \\ of spk}} & \textbf{Domain} & \textbf{Condition} \\
    \midrule
    CV \cite{commonvoice} & 7.7 & 100 & read. & regular voice \\
    LS test-clean & 2.2 & 40 & read. & regular voice \\
    Chains\tablefootnote{https://chains.ucd.ie} & 2.4 & 36 & read. & \multirow{3}{*}{\hspace{-1em}$\left.\begin{array}{l}
                \\
                \\
                \\
                \end{array}\right\lbrace$ \makecell{ regular \\ voice, \\ whisper}} \\ 
    WhiSp & 4.4 & 40 & spont. \\
  \end{tabular}
\end{table}

% Based on this test datasets, we consider variety of testing protocols, which can be categorized into four different types. \textbf{Whisper-to-speech} condition, where test audio is initially whispered and then converted to speech of same speaker. \textbf{Cross-speaker whisper-to-speech} condition, where test audio is also initially whispered, but then converted to the voice of another speaker. \textbf{Cross-speaker speech-to-speech} condition is built according to same logic, but, test audio is initially voiced. The enrollment audio is voiced in all described conditions.
%It must be noted that the Chains dataset contains parallel recordings of whispered and voiced data, where the same speakers read the same sentence in both whispered and regular voice.

\section{Systems description}

 %Inspired by the approach in \cite{TTS_yourTTS, VC_cycle_cons}, we also consider incorporating additional speaker loss during training for all our systems. 
In this section, we describe the considered VC systems. Inspired by the approach outlined in \cite{TTS_yourTTS, VC_cycle_cons}, we employ cosine Speaker Loss (SL) during the training of the decoders. We also describe the proposed SpeakerVC system, which is based on the StyleTTS2 approach \cite{TTS_style_tts} adapted for the VC task.

\subsection{Encoder}
Our system is based on the HuBERT \cite{hubert} encoder with adaptation to the streaming condition and whispered domain as described in \cite{our_icassp}. We also modify this model with the approach from the HuBERT-Soft article \cite{hubert_soft} to obtain soft speech units for decoders. To learn discrete speech units, we apply k-means clustering with 1024 clusters. Then, we train a linear projection layer between the discrete features and the backbone network without updating the backbone network weights to keep the phonetic information unchanged. We use the output of the projection layer to extract soft speech units, which are used as input features for all our decoders. 
%We utilize phonetic posteriorgrams (PPGs) extracted from the HuBERT-Soft model \cite{hubert_soft} as the input acoustic features for all our systems. Additionally, we adapt this model to the streaming  condition and whispered domain as described in [to ensure author anonymity, the link to the resource will be added after the review process]. 

\subsection{Speaker loss}

For all our experiments with SL we use ECAPA-TDNN model \cite{ecapa_tdnn} as a speaker encoder. The idea behind the additional loss function is simply to minimize the distance between embeddings extracted from the reference and generated with the VC system audio. For this, we use Cosine Loss: 
\begin{equation}
    \mathcal{L}_{spk} = \frac{1}{N}\left( 1 - \frac{X \cdot Y}{\lVert X \rVert \lVert Y \rVert} \right)
\end{equation}
where, $X$ represents embeddings extracted from reference speech, $Y$ represents embeddings extracted from generated speech, $N$ -- batch size. Our experiments with more complex loss functions, such as AMSoftmax \cite{am_softmax}, did not show a significant increase in speaker similarity metrics compared to the cosine loss function. 

\subsection{Tacotron-based decoder} \label{subsection:tacotron}

Our first decoder is based on the Tacotron 2 model adapted to the streaming condition as described in \cite{our_icassp}. To speed-up computation during training we consider additional Mel adaptor in case of using SL. Since the Tacotron 2 model and the ECAPA-TDNN model have different feature extraction parameters, the aim of the Mel adaptor is to convert mel spectrograms extracted with one set of settings to mel spectrograms extracted with another set of settings. Thus, a 3-layer TDNN model was pretrained for 20 epochs on the VCTK dataset to implement such feature conversion. The Mel adaptor block remains frozen during decoder training to avoid model overfitting under the SL minimisation task. As a vocoder we use a pretrained HI-FI GAN model \cite{hifi_gan}.

\begin{figure}[t]
  \centering
  \includegraphics[width=1.1\linewidth, height=0.6\linewidth]{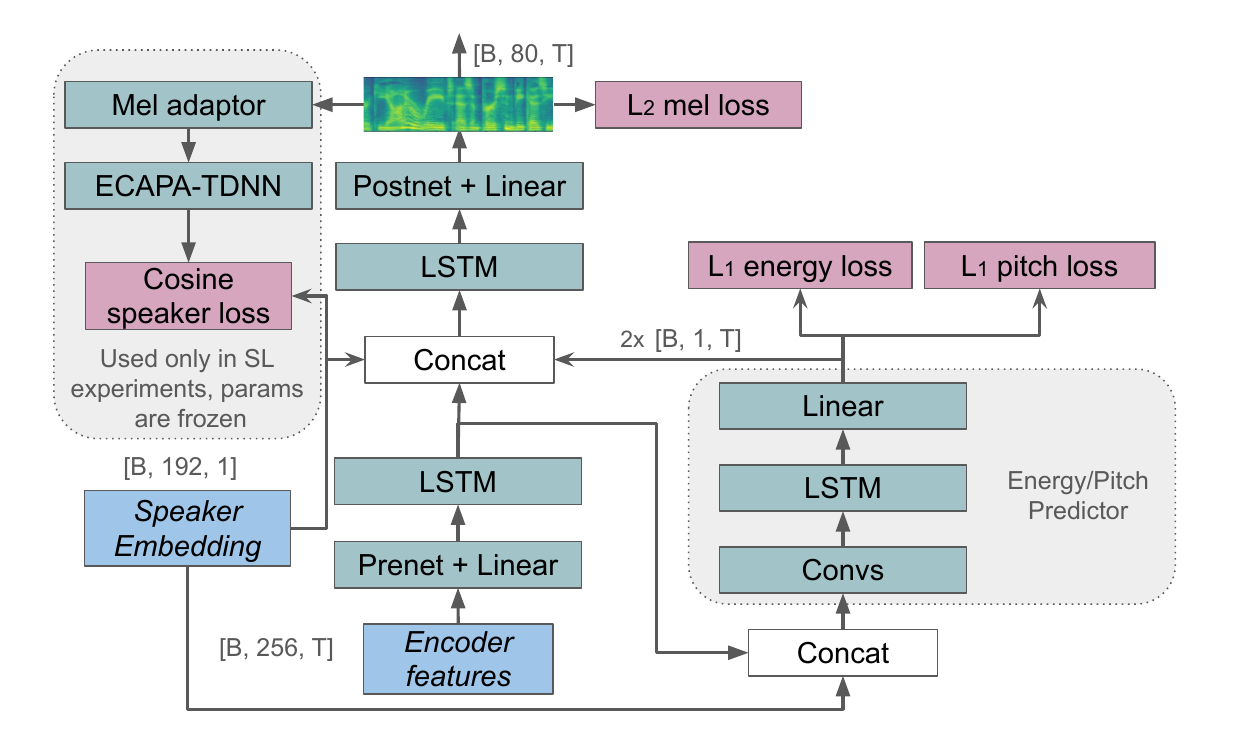}
  \caption{Proposed FastSpeech2-based VC system.}
  \label{fig:FastSpeech2}
\end{figure}

\subsection{FastSpeech-based decoder}
As an alternative approach we use the FastSpeech 2 model \cite{Fastspeech2} with modifications to the base architecture for the VC task. %Unlike our previous experiments with Tacotron-based decoder, we found Energy and Pitch Predictor blocks useful to improve the quality of whisper-to-speech conversion by this system. 
We change the Transformer block to LSTM in the Encoder and Decoder blocks of FastSpeech 2 model and add LSTM layers in the Energy and Pitch Predictor blocks as a trade-off between better memorizing of past context and adaptation to chunk-wise processing. We remove the default speaker embedding layer and use the ECAPA-TDNN model \cite{ecapa_tdnn} to extract speaker embeddings. As for the Tacotron-based model, we also utilize the Mel adaptor block in case of using SL. As a vocoder we use a pretrained HI-FI GAN model. Overall, the proposed system is shown in Figure~\ref{fig:FastSpeech2}.

\subsection{SpeakerVC}
\begin{figure}[t]
  \centering
  \includegraphics[width=1.1\linewidth, height=0.6\linewidth]{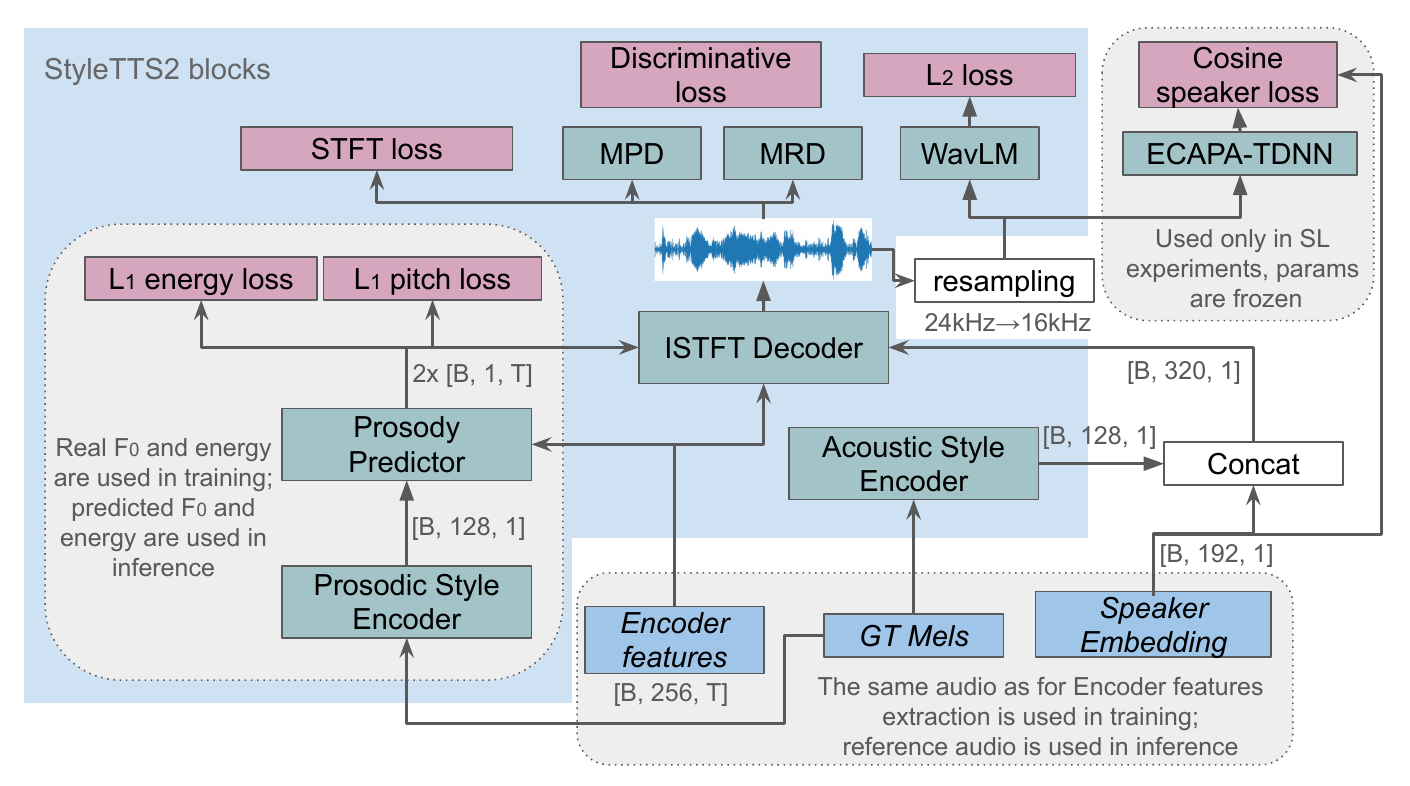}
  \caption{Proposed SpeakerVC system.}
  \label{fig:StyleVC}
\end{figure}

The third system we propose is based on the StyleTTS2 model \cite{TTS_style_tts} with modifications of the architecture for the VC task. We use speaker embeddings from the ECAPA-TDNN model in addition to the Acoustic Style Encoder to encode speaker information. The Prosody Predictor and Prosody Style Encoder blocks are trained independently of the rest of the model. Also, to simplify model training, we discarded the Style Diffusion block and the Speech Language model based discriminator block. Several changes have also been made to the training process from the base paper. As an input for the Acoustic Style Encoder we use the same audio fragment as for the ground true. Waveforms are randomly cropped with a maximum length of 4 seconds. We change the hop length mel spectrogram parameter to 240 to better synchronise the time resolution between features at the decoder input and encoder output. Additionally, we consider a third stage of training in which the SL is added to the initial losses to fine-tune the decoder, aiming to achieve better speaker similarity. Overall, the proposed system is shown in Figure~\ref{fig:StyleVC}.

\begin{table*}[t]
  \caption{Performance of our various VC systems in terms of EER metric (\%) and SIM-o metric. Protocols denoted with "*" assume that test samples are initially whispered, and voiced enrollments are compared against the results of whisper-to-speech VC. The Chains2WhiSp and LS2CV protocols correspond to the cross-speaker and cross-dataset protocol, meaning that, for example, in the case of the Chains2WhiSp protocol, the whispered speech of each speaker from the Chains dataset is converted to the voice of a speaker from the WhiSp dataset.}
  \label{tab:eer_comp}
  \centering
  \begin{tabular}{l|*{10}c}
    \toprule
    \textbf{model} & \multicolumn{2}{c}{\textbf{Chains*}} & \multicolumn{2}{c}{\textbf{WhiSp*}} & \multicolumn{2}{c}{\textbf{Chains2WhiSp*}} & \multicolumn{2}{c}{\textbf{CV}} & \multicolumn{2}{c}{\textbf{LS2CV}} \\
    & EER & SIM-o & EER & SIM-o & EER & SIM-o & EER & SIM-o & EER & SIM-o \\
    \midrule
    speech-vs-speech (GT) & 0.189 & 0.788 & - & - & -- & -- & 0.939 & 0.756 & -- & -- \\
    speech-vs-whisper & 9.396 & 0.444 & 9.467 & 0.463 & -- & -- & -- & -- & -- & -- \\
    \midrule
    Tacotron-based & 6.666 & 0.470 & 5.890 & 0.457 & 4.401 & 0.367 & 15.175 & 0.316 & 14.267 & 0.328  \\
    + SL & 1.386 & 0.583 & 1.258 & 0.560 & 1.077 & 0.460 & 4.251 & 0.414 & 5.374 & 0.431 \\
    + SL + D\_EXT & 0.332 & 0.620 & 0.946 & 0.615 & 0.947 & 0.525 & 1.355 & 0.514 & 1.584 & 0.504 \\
    \midrule 
    FastSpeech-based & 5.261 & 0.496 & 3.971 & 0.475 & 4.085 & 0.380 & 9.136 & 0.344 & 6.916 & 0.345 \\
    + SL & 0.458 & 0.585 & 1.240 & 0.596 & 1.281 & 0.510 & 4.062 & 0.455 & 2.773 & 0.455 \\
    + SL + D\_EXT & 0.287 & 0.637 & 1.127 & 0.615 & 0.925 & 0.539 & 1.115 & 0.528 & 1.386 & 0.522 \\
    \midrule
    SpeakerVC & 10.741 & 0.321 & 11.941 & 0.324 & 10.436 & 0.206 & 16.41 & 0.181 & 15.888 & 0.199 \\
    + Spk. Emb. & 1.557 & 0.539 & 1.165 & 0.543 & 1.238 & 0.444 & 6.865 & 0.370 & 7.140 & 0.383 \\
    + Spk. Emb. + SL & 0.406 & 0.609 & 1.156 & 0.623 & 1.034 & 0.510 & 1.760 & 0.481 & 6.130 & 0.445 \\
    + Spk. Emb. + SL + D\_EXT & \textbf{0.021} & \textbf{0.733} & \textbf{0.150} & \textbf{0.741} &  \textbf{0.109} & \textbf{0.646} & \textbf{0.819} & \textbf{0.643} & \textbf{1.195} & \textbf{0.631} \\
  \end{tabular}
\end{table*}

\subsection{Evaluation metrics}
For evaluating the results, we utilize both objective and subjective metrics. As a subjective evaluation metric we use the Similarity Mean Opinion Score (SMOS), estimated through the Toloka platform\footnote{https://toloka.ai}. We select 20 speakers from each of the WhiSp and LS test-clean datasets. For each speaker, we perform voice conversion using our various proposed systems and the Pheme VC system, utilizing another sample of the same speaker's voice as the enrolled speech. Subsequently, 30 Toloka users were asked to evaluate the similarity between each generated and the real sample of the speaker's voice on the 1-5 scale. The used scale is similar to that in \cite{mos_settings}. As an objective evaluation metrics, Word Error Rate (WER), Equal Error Rate (EER) and speaker similarity are employed to measure quality of content and speaker reconstruction. We utilize the speaker similarity measure based on the WavLM-TDNN model \cite{wavlm_tdnn}, as proposed in \cite{voicebox}. The HuBERT-Large model \cite{hubert} is used in the WER evaluation.
%Additionally, we utilize the speaker similarity measure based on the WavLM-TDNN model \cite{wavlm_tdnn}, as proposed in \cite{voicebox}.
%For WER evaluation, the Conformer-Transducer-Large model for English ASR (Conformer-T-L) \footnote{https://catalog.ngc.nvidia.com/orgs/nvidia/teams/nemo/models\\/stt\_en\_conformer\_transducer\_large} is used.
For the EER evaluation, we employ variety of verification protocols, which can be categorized into three different types. \textbf{Whisper-to-speech} condition, where test audio is initially whispered and then converted to speech of same speaker. \textbf{Cross-speaker whisper-to-speech} condition, where test audio is also initially whispered, but then converted to the voice of another speaker. \textbf{Cross-speaker speech-to-speech} condition is built according to same logic, but, test audio is initially voiced. The enrollment audio is voiced in all the described conditions. We use the Nemo TitaNet-L\footnote{https://catalog.ngc.nvidia.com/orgs/nvidia/teams/nemo/models/titanet\_large} model to extract speaker embeddings and obtain scores for the EER calculation.
% We use Nemo TitaNet-L\footnote{https://catalog.ngc.nvidia.com/orgs/nvidia/teams/nemo/models/titanet\_large} model to extract speaker embeddings and obtain scores for EER calculation. 

We also consider comparing of our best performing systems with the SOTA zero-shot TTS and VC systems. For this evaluation we employ \textbf{Sim-o} test settings, as proposed in \cite{voicebox}, and compute similarity against original audio. Following the evaluation procedures from \cite{vall_e, voicebox}, we select files ranging from 4 to 10 seconds in length from the LS test-clean dataset and consider only \textit{cross-sentence} type of test conditions. We utilize a 3-second clip from another sample of the same speaker for style and speaker embedding extraction. Additionally, we also employ a \textit{cross-speaker} condition, which is more related to the VC task. In this scenario, a 3-second clip from a random sample of another speaker is used for style and embedding extraction, representing \textbf{any-to-any VC}. 
%For the intelligibility testing the HuBERT-L model \cite{hubert} is used in WER evaluation. 

\begin{table}[h]
  %\caption{Results on filtered LS test-clean. We use chunked processing for SpeakerVC streaming with a total delay of 0.8s. and a past context of 2.0s. Voicebox and Vall-e results are taken from \cite{voicebox}. Pheme \footnote{https://github.com/PolyAI-LDN/pheme} and StyleTTS2 \footnote{https://github.com/yl4579/StyleTTS2/tree/main} results are obtained using a publicly available models. SpeakerVC adapt and FastSpeach-based adapt models correspond to our best performing models according to results in Table \ref{tab:eer_comp}. For the intelligibility testing, following \cite{vall_e}, the HuBERT-L model \cite{hubert} is used for WER evaluation.}
    \caption{Results on the filtered LS test-clean dataset. We use chunked processing for SpeakerVC streaming with a total delay of 0.8s. and a past context of 2.0s. YourTTS and Voicebox results are taken from \cite{voicebox}. Pheme and StyleTTS2 results were obtained using publicly available models. SpeakerVC adapt and FastSpeech-based adapt models correspond to our best performing models according to the results in Table \ref{tab:eer_comp}.}
  \label{tab:sota_comparing}
  \centering
  \begin{tabular}{l|cc}
    \toprule
    \textbf{Model} & \textbf{WER} & \textbf{Sim-o} \\
    \midrule 
    Ground truth & 2.2 & 0.754 \\
    \midrule
    \textit{cross-sentence} \\
    % Vall-E \cite{vall_e} & 5.9 & - \\
    YourTTS \cite{TTS_yourTTS} & 7.7 & 0.337 \\
    Voicebox \cite{voicebox} & \textbf{1.9} & 0.662 \\
    Pheme \cite{pheme} & 4.9 & 0.466 \\
    StyleTTS2 \cite{TTS_style_tts} & 2.6 & 0.335 \\
    SpeakerVC adapt. & 2.6 & \textbf{0.710} \\
    SpeakerVC adapt. streaming & 3.1 & 0.694 \\
    FastSpeech-based adapt. & 2.7 & 0.633 \\
    \midrule
    %\textit{continuation} \\
    %Vall-E \cite{vall_e} & 3.8 & %0.452 \\
    %\midrule
    % Voicebox \cite{voicebox} & 2.0 & 0.593 \\
    % StyleVC + ECAPA + D\_ext & 2.42 & 0.655 \\
    % Parallel Decoder + SL + D\_ext  & & \\
    \textit{cross-speaker} \\
    Pheme & 4.9 & 0.372 \\
    SpeakerVC adapt. & \textbf{2.6} & \textbf{0.689} \\
    SpeakerVC adapt. streaming & 3.4 & 0.674 \\
    FastSpeech-based adapt. & 2.8 & 0.607 \\
  \end{tabular}
\end{table}

\section{Experiments}
Table \ref{tab:eer_comp} presents our main results of speaker similarity evaluation. \textit{Chains} and \textit{WhiSp} here correspond to whisper-to-speech condition, \textit{Chains2WhiSp} corresponds to cross-speaker whisper-to-speech condition, while \textit{CV} and \textit{LS2CV} correspond to cross-speaker speech-to-speech conditions. The first two rows of Table \ref{tab:eer_comp} demonstrate the EER evaluation on the original data before voice conversion. For whispered data, we consider EER evaluation between voiced enrollment and whispered test. Parallel whispered and voiced recordings from Chains allow us to compare speaker similarity in both conditions. 
%It can be observed that the quality of speaker embedding models significantly degrades in the whispered domain. 
According to the achieved results, it can be seen that the use of SL significantly improves the speaker similarity between the original audio and its conversion and reduces the EER on test datasets for different VC systems. Additionally, increasing the number of speakers in the training dataset leads to further quality improvement for for all considered VC systems. For the SpeakerVC system, employing ECAPA-TDNN embedding along with the Acoustic Style Encoder leads to a significant improvement in terms of EER and the speaker similarity metric. 
%Both using speaker loss and increasing the diversity of the train dataset, these effects are stronger for more recent VC models. 
%Note also that for some of speech-to-speech protocols, EER value after conversion even lower than that obtained from the original data.
%Note also that in some cases the final EER metric on test datasets, where enroll is the original regular speech and test is the converted regular speech, is lower than for datasets consisting entirely of regular speech. 

%Table \ref{tab:sota_comparing} presents the results of comparing the proposed system with various SOTA systems in terms of speaker similarity measure. 
%This includes our systems, Voicebox \cite{voicebox}, and other models evaluated on the LS test-clean dataset. 
%As found, the proposed SpeakerVC system outperforms existing SOTA TTS solutions such as Voicebox, as well as Pheme system in VC mode in terms of speaker similarity metric. 
Table \ref{tab:sota_comparing} presents the results of comparing the proposed systems with various SOTA systems in terms of speaker similarity measure, reveals that the proposed SpeakerVC system outperforms existing TTS solutions such as Voicebox, as well as Pheme system in VC mode.
In the cross-speaker condition, we compare the proposed systems with the Pheme system. It can be observed that both SpeakerVC and FastSpeech-based systems show superior speaker similarity quality and only slightly degrade when moving from cross-sentence to cross-speaker condition. For SpeakerVC it has been also shown that the model is robust in streaming processing.

%Table \ref{tab:wer_mos} demonstrates the results of the MOS evaluation. As found, the results of subjective evaluation are mostly correlate with the results of objective evaluation. 
Table \ref{tab:wer_mos} displays the SMOS evaluation results, revealing a notable correlation between the subjective and objective evaluation outcomes. However, the improvement obtained from utilizing the SL is not as significant in SMOS evaluation as it is in terms of the speaker similarity metric. Furthermore, the performance of SpeakerVC adapted system is comparable to Pheme in the regular voice domain and better in whisper domain, for which the system was fine-tuned. Moreover, the proposed SpeakerVC system is almost two times faster.

%\begin{table}[h]
%  \caption{Results of WER and MOS evaluation. MOS results are given with 95\% confidence interval.}
%  \label{tab:wer_mos}
%  \centering
%  \begin{tabular}{l|ccсс}
%    \toprule
%    \textbf{Model} & \makecell{\textbf{WER} \\ WhiSp} & \makecell{\textbf{MOS} \\ WhiSp} & \makecell{\textbf{WER} \\ LS} & \makecell{\textbf{MOS} \\ LS } \\
%    \midrule
%    Pheme & 70.0 & & 6.1 & \\
%    SpeakerVC & 10.5 & & 2.6 & \\
%    SpeakerVC adapt. & 17.3 & & 2.6 & \\
%    FastSpeech-based & 11.6 & & 2.9 &\\
%    FastSpeech-based adapt. & 13.2 & & 2.9 & \\
%  \end{tabular}
%\end{table}

\begin{table}[h]
  \caption{Results of the SMOS evaluation. SMOS results are given with 95\% confidence interval. Pheme results were obtained using a publicly available model. For the GT another sample of the same speaker is used.}
  %\caption{Results of MOS evaluation. MOS results are given with 95\% confidence interval. Pheme results were obtained using a publicly available model. For the GT another sample of the same speaker is used. RTF results are obtained on an Nvidia 4090 GPU for the full VC pipeline.}
  \label{tab:wer_mos}
  \centering
  \begin{tabular}{l|ccc}
    \toprule
    \textbf{Model} & \makecell{ \textbf{SMOS} \\ WhiSp} & \makecell{\textbf{SMOS} \\ LS} & \textbf{RTF} \\
    \midrule
    GT & 4.75 $\pm$ \scalebox{0.8}{0.02} & 4.54 $\pm$ \scalebox{0.8}{0.07} & -- \\
    \midrule
    Pheme & 2.31 $\pm$ \scalebox{0.8}{0.11} & 3.40 $\pm$ \scalebox{0.8}{0.17} & 0.055 \\
    SpeakerVC & 3.61 $\pm$ \scalebox{0.8}{0.12} & 3.20 $\pm$ \scalebox{0.8}{0.18} &  \textbf{0.029} \\
    SpeakerVC adapt. & \textbf{3.75} $\pm$ \scalebox{0.8}{0.11} & \textbf{3.46} $\pm$ \scalebox{0.8}{0.18} & 0.031 \\
    FastSpeech-based & 2.96 $\pm$ \scalebox{0.8}{0.12} & 2.97 $\pm$ \scalebox{0.8}{0.18} & 0.033 \\
    FastSpeech-based adapt. & 3.05 $\pm$ \scalebox{0.8}{0.12} & 3.08 $\pm$ \scalebox{0.8}{0.17} & 0.033 \\
  \end{tabular}
\end{table}

\section{Discussion}
In this paper, we proposed the SpeakerVC model -- a fast, streamable and robust system for zero-shot any-to-any whispered and regular speech VC. We also considered methods to improve the speaker similarity of the converted speech and demonstrated their effectiveness across various VC systems. Despite the significant improvement in speaker similarity transfer achieved by utilizing the speaker loss during training and increasing the number of speakers in the training dataset, some issues remain unresolved. According to the SMOS results, there is still a gap between generated and real speech. Also, there is mismatch between objective and subjective evaluations. While differences between systems are significant in terms of objective metrics such as EER and speaker similarity, these systems show close quality in SMOS evaluation. 

\bibliographystyle{IEEEtran}
\bibliography{main}

\end{document}